\documentclass[12pt,preprint2]{emulateapj}
\usepackage{natbib}
\usepackage{amsmath}
\usepackage{graphicx}
\usepackage{wasysym}
\usepackage{txfonts}
\usepackage{xspace}
\usepackage{url}
\usepackage{textcomp}
\usepackage{multirow}
\usepackage{lipsum}

\newcommand{\gx}{GX~339$-$4\xspace}

\newcommand{\swift}{\textsl{Swift}\xspace}

\newcommand{\suz}{\textsl{Suzaku}\xspace}

\newcommand{\xte}{\textsl{RXTE}\xspace}

\newcommand{\nustar}{\textsl{NuSTAR}\xspace}

\newcommand{\snr}{S/N\xspace}

\newcommand{\msun}{\ensuremath{\text{M}_{\odot}}\xspace}

\newcommand{\redchi}{\ensuremath{\chi^{2}_\text{red}}\xspace}

\newcommand{\feka}{\ensuremath{\mathrm{Fe}~\mathrm{K}\alpha}\xspace}

\newcommand{\ledd}{\ensuremath{{L}_{\mathrm{Edd}}}\xspace}
\renewcommand{\deg}{\ensuremath{^\circ}\xspace}
\newcommand{\asec}{\ensuremath{''}\xspace}
\newcommand{\nuqpo}{\ensuremath{\nu_\text{QPO}}\xspace}

\shorttitle{Spectro-timing of \gx in an HIMS}
\shortauthors{F\"urst et al.}

\begin{document}

\title{Spectro-timing study of  \gx  in a hard intermediate  state}

\author{F.~F\"urst\altaffilmark{1}}
\author{V.~Grinberg\altaffilmark{2}}
\author{J.~A.~Tomsick\altaffilmark{3}}

\author{M.~Bachetti\altaffilmark{4}}
\author{S.~E.~Boggs\altaffilmark{3}}
\author{M.~Brightman\altaffilmark{1}}
\author{F.~E.~Christensen\altaffilmark{5}}
\author{W.~W.~Craig\altaffilmark{3,6}}
\author{P.~Gandhi\altaffilmark{7}}
\author{B.~Grefenstette\altaffilmark{1}}
\author{C.~J.~Hailey\altaffilmark{8}}
\author{F.~A.~Harrison\altaffilmark{1}}
\author{K.~K.~Madsen\altaffilmark{1}}
\author{M.~L.~Parker\altaffilmark{9}}
\author{K.~Pottschmidt\altaffilmark{10,11}}
\author{D.~Stern\altaffilmark{12}}
\author{D.~J.~Walton\altaffilmark{12,1}}
\author{J.~Wilms\altaffilmark{13}}
\author{W.~W.~Zhang\altaffilmark{11}}



\altaffiltext{1}{Cahill Center for Astronomy and Astrophysics, California Institute of Technology, Pasadena, CA 91125, USA}
\altaffiltext{2}{Massachusetts Institute of Technology, Kavli Institute for Astrophysics, Cambridge, MA 02139, USA}
\altaffiltext{3}{Space Sciences Laboratory, University of California, Berkeley, CA 94720, USA}
\altaffiltext{4}{INAF/Osservatorio Astronomico di Cagliari,  09047 Selargius (CA), Italy} 
\altaffiltext{5}{DTU Space, National Space Institute, Technical University of Denmark, 2800 Lyngby, Denmark} 
\altaffiltext{6}{Lawrence Livermore National Laboratory, Livermore, CA 94550, USA}
\altaffiltext{7}{Department of Physics and Astronomy, University of Southampton, Highfield, Southampton SO17 1BJ, United Kingdom}
\altaffiltext{8}{Columbia Astrophysics Laboratory, Columbia University, New York, NY 10027, USA}
\altaffiltext{9}{Institute of Astronomy,  Cambridge CB3 0HA, UK} 
\altaffiltext{10}{CRESST, Department of Physics, and Center for Space Science and
Technology, UMBC, Baltimore, MD 21250, USA}
\altaffiltext{11}{NASA Goddard Space Flight Center, Greenbelt, MD 20771, USA}
\altaffiltext{12}{Jet Propulsion Laboratory, California Institute of Technology, Pasadena, CA 91109, USA}
\altaffiltext{13}{Dr. Karl-Remeis-Sternwarte and ECAP,  University of Erlangen-Nuremberg, 96049 Bamberg, Germany} 


\begin{abstract}
We present an analysis of \nustar observations of a hard intermediate state of the transient black hole \gx taken in January 2015. As the source softened significantly over the course of the 1.3\,d-long observation we split the data into 21 sub-sets and find that the spectrum of all of them can be well described by a power-law continuum with an additional relativistically blurred reflection component. The photon index increases from $\sim$1.69 to $\sim$1.77 over the course of the observation. The accretion disk is truncated at around 9 gravitational radii in all spectra. We also perform timing analysis on the same 21 individual data sets, and find a strong type-C quasi-periodic oscillation (QPO), which increase in frequency from $\sim$0.68 to $\sim$1.05\,Hz with time. The frequency change is well correlated with the softening of the spectrum. We discuss possible scenarios for the production of the QPO and calculate predicted inner radii in the relativistic precession model as well as the global disk mode oscillations model. We find discrepancies with respect to the observed values in both models unless we allow for a black hole mass of $\sim$100\,\msun, which is highly unlikely. We discuss possible systematic uncertainties, in particular with the measurement of the inner accretion disk radius in the relativistic reflection model. We conclude that the combination of observed QPO frequencies and inner accretion disk radii, as obtained from spectral fitting, is difficult to reconcile with current models.
\end{abstract}


\section{Introduction}
\label{sec:intro}

While we have gathered a wealth of information about accretion physics in black hole binaries in the past decades, new and improved data still uncover new phenomena. Most of the known accreting black holes are transient, and undergo different spectral states over the course of a typical outburst. The two most important states are the hard state, in which a power-law with a photon index $\Gamma \lesssim 1.8$ dominates the X-ray spectrum, and the soft state, in which a soft  thermal component is dominant and the power-law is steeper. Superimposed on these continua are reflection features from radiation reprocessed within the accretion disk, namely the \feka line around 6.4\,keV and a Compton hump between 10--30\,keV. These features are also subject to strong relativistic effects close to the black hole.
Typical outbursts start in the hard state and switch to the soft state at luminosities $\sim10\%$ of the Eddington luminosity (\ledd) or above. 

The soft  component of the spectrum can be identified with thermal emission from the optically thick, geometrically thin accretion disk \citep{shakura73a}. This disk provides a steady flow of material to the black hole, which is confirmed by the lack of variability during the soft state, where it is dominant \citep[see, e.g.,][]{belloni10a}. In the hard state, on the other hand, a Comptonizing electron gas dominates the X-ray spectrum (the so-called corona) and a very high RMS variability is observed. The location and physics of the corona are still unknown and with it the extent of the accretion disk. 

During the hard state, the accretion flow is unstable, i.e., shows large aperiodic variability, and the mass accretion rate is reduced. It is likely that the inner parts of the disk are replaced by an advection-dominated accretion-flow \citep[ADAF,][]{narayan95a, esin97a}, or that  the inner accretion disk is otherwise truncated outside the innermost stable circular orbit (ISCO). 
 Observations  clearly indicate that at the lowest luminosities the accretion disk is indeed truncated at several 10s to 100 $r_g$ \citep{tomsick09a, allured13a}.  At higher luminosities ($\geq1\%\,\ledd$) most data require an inner radius very close to the ISCO  \citep[see, e.g.][]{nowak02a, miller06a, petrucci14a, miller15a}, although other authors found that the disk is truncated throughout the hard state \citep[e.g.][]{done10a, plant15a} or that truncation may not be a linear function of luminosity \citep{kolehmainen14a}. Therefore the question of when the disk moves towards the ISCO remains open: is it during the brighter phases of the hard state or later during the state transition?

A further complication  is that not all outbursts  follow the standard evolution. During  so-called failed outbursts, the source never leaves the hard state, and declines in flux before the switch to the soft state is achieved. Such an outburst of \gx was observed in detail with \nustar in 2013 \citep[hereafter F15]{gx339}. Using a combination of \swift/XRT and \nustar spectra we could show that the 0.8--75\,keV spectrum exhibits a more complex reflection geometry than previously assumed, in which the reflector sees a significantly harder spectrum than the observed primary continuum. Due to that complex geometry, measurements of the inner radius of the accretion disk were challenging and dominated by systematic uncertainties of the applied model.

If the source goes through a full outburst, the change from the hard state to the soft state and vice versa happens rather quickly, on time scales of days (compared to typical outburst durations of months). These transitions show a hysteretic behavior, i.e., the hard-to-soft transition typically happens at higher luminosities than the switch back to the hard state at the end of the outburst. During the transitions, the energy and power spectra  often show peculiar qualities.  These intermediate states were first described by \citet{miyamoto91a} and \citet{mendez97a}, among others, although at that time a distinction was made between an intermediate state at low luminosities during the decline of the outburst and the very high state at higher luminosities during the rise. Later works recognized the similarity between these two states  \citep[see, e.g.][]{belloni97a, mcclintock03a}, and \citet{homan05a} introduced new classifications based on the spectral hardness and timing properties,  the so-called hard intermediate state (HIMS) and  soft intermediate state (SIMS). Both these states show similar energy spectra, in which a thermal component is detected but the power-law continuum dominates the flux. Their timing properties, however, differ \citep[see, e.g.,][]{rutledge99a, boeck11a}: while the HIMS can be described with band-limited noise components and a narrow QPO similar to the hard state, the SIMS power spectra are dominated by a type-B QPO. The contrast in the shape of the power spectra is very stark between these two states \citep{belloni05a, belloni06a}.

These intermediate states are difficult to observe due to their short duration. They were first observed and described in detail in the prototypical X-ray black hole binary \gx, which shows frequent outbursts and has therefore been regularly observed since its discovery \citep{markert73a}.
The most detailed study of a  HIMS in \gx to date was presented by \citet{tamura12a}. These authors studied \suz data taken during a state transition in 2007 and find that the broad-band spectrum (0.7--200\,keV) is well described by a combination of a thermal accretion disk and a power-law continuum, with a photon index of $\Gamma\approx2.68$. They find indications that the accretion disk is truncated marginally outside  the ISCO, which  would indicate that the disk only moves all the way to the ISCO  at the end of the transition. The QPO frequency and behavior observed by \citet{tamura12a} is in agreement with predictions from the relativistic precession model (RPM), in particular with Lense-Thirring precession of the inner accretion flow \citep[see, e.g.,][]{ingram09a}.

\gx  provides an ideal target to be studied with the \textsl{Nuclear Spectroscopic Telescope Array} \citep[\nustar,][]{harrison13a}, which allows for very sensitive measurements of the accretion geometry and the inner accretion disk radius.
While the X-ray and radio properties of \gx are very well studied, its companion and therefore its orbital ephemeris is still unknown. Best estimates of the black hole mass and distance are $M=9.0^{+1.6}_{-1.2}\,\msun$ and $8.4\pm0.9$\,kpc, respectively \citep{parker16a}, which we will adopt here. These are close to typically assumed values of $M=10\,\msun$ and  $d$=10\,kpc,
See \citet{parker16a} for a discussion on the impact of this choice.

The remainder of the paper is structured as follows: in Section~\ref{sec:data} we described the data  and detail the extraction method of the energy spectra and power  spectral density (PSD). The main analysis of the light curve, the energy spectra and PSD is described in Section~\ref{sec:analysis}. In Section~\ref{sec:summ} we summarize and discuss our results.

\section{Observations and data reduction}
\label{sec:data}

\begin{figure}
\includegraphics[width=0.95\columnwidth]{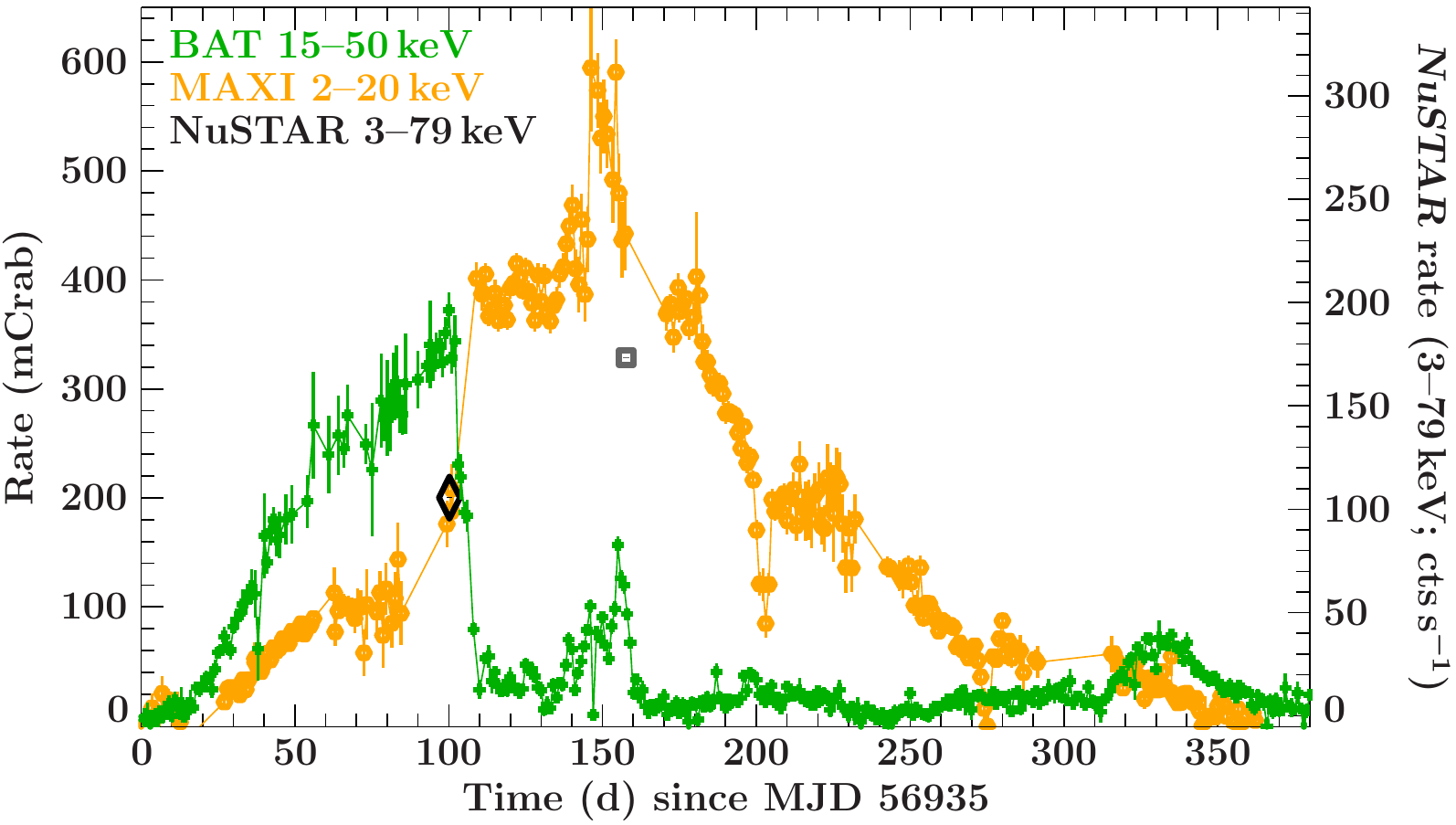}
\caption{\swift/BAT \citep[15--50\,keV, green;][]{swiftbatref} and MAXI \citep[2--10\,keV, orange; ][]{maxiref} monitoring light curve of the 2015 outburst of \gx. The \nustar observation (3--79\,keV) presented here is marked by a black diamond, the one used by \citet{parker16a} by a square. All count-rates have been rescaled to mCrab fluxes in the respective energy band of the instrument. The right-hand $y$-axis gives the average  measured \nustar count-rate of the observation. The dip seen around MJD\,57135 in the MAXI light curve is an instrumental effect (T. Belloni, priv. comm.). }
\label{fig:batlc}
\end{figure}

In late 2014, \gx showed renewed activity (see Figure~\ref{fig:batlc}), starting as expected in a low hard state \citep{yan14a}. Around MJD 57035 the source switched into the soft state \citep{motta15a}, and faded into quiescence about 350\,d after the beginning of the outburst. We obtained a \nustar observation just as the source was transitioning and found it in  an HIMS. A second \nustar observation was performed about 50\,d later during a peak in the soft state, which is presented by \citet{parker16a}. In this article we concentrate on the intermediate state observation and compare its spectral behavior to the results of F15.

\nustar observed \gx on 2015 January 13 and 14 (MJD 57035.24--57036.64) for a total exposure time of 55\,ks per module (ObsID 80001015001). At that time, \gx was very close to the sun and not observable by \swift/XRT or any other soft X-ray telescope. 

The small Solar aspect angle also influenced \nustar's optical bench star tracker which is co-aligned with the X-ray optics. The tracker's camera was saturated for about 80\% of the time, during which the science data are stored in mode 06 (SCIENCE\_SC). The aspect reconstruction during mode 06 uses the spacecraft bus star trackers instead of the co-aligned star tracker, resulting in larger positional uncertainties. Consequently the source was not reconstructed to a perfect point source on the sky, but shifts by about 40\asec per satellite orbit. Thanks to the laser metrology system the location of the optical axis on the detector is still known, allowing for an extraction of the spectroscopic and timing information  and the construction of correct response files. See Walton et al. (2016, subm.) for a detailed description of the mode 06 extraction procedure.

\subsection{Energy spectra}

We extracted mode 01 (SCIENCE, $\sim$10\,ks per module) and mode 06 (SCIENCE\_SC, $\sim$45\,ksec per module) data separately using a 90\asec radius extraction region centered at the J2000 coordinates of \gx for the mode 01 data and a larger 120\asec radius region for the mode 06 data. The larger region for mode 06 data was chosen to encompass the whole source at all times, as the region was defined in sky coordinates. We carefully compared the extracted spectra and found no significant differences in spectral shape or response. We then combined the data from the two modes using \texttt{addascaspec}, separately for focal plane modules (FPM) A and B.

For time-resolved spectroscopy we used manually created good-time interval (GTI) files and supplied them to the pipeline and followed the same procedure as describe above. 
The fastest time-scales we extracted were of the length of one revolution of the \nustar satellite, corresponding to an average exposure of $\sim 2.7$\,ks per spectrum.

To measure the background  we extracted spectra from a 120\asec region north of the source at the other end of the field of view. This places the region on a different detector with slightly different intrinsic background, but \gx dominates over the background by a factor of at least 10 up to  78\,keV. Small background fluctuations  therefore do not influence our results.

Data were extracted using \texttt{nupipeline} v.1.4.1 as distributed with HEASOFT v6.16 and CALDB files v20150316. At the time of writing, data below 4.5\,keV suffer from slightly higher calibration uncertainties, due to a known shift in gain offset that has not been properly accounted for in the current response files. They have therefore been ignored in the spectral analysis but are still reliable for the light curves. As the relevant features for our conclusions (iron line, photon index) are located well above 4.5\,keV, we do not expect this limitation to influence our results significantly.
All analysis was performed using the Interactive Spectral Interpretation System \citep[ISIS v1.6.2,][]{houck00a} and uncertainties are reported at the 90\% confidence level unless otherwise noted.

\subsection{Timing data}
We transferred the photon arrival times to the barycenter of the Solar System using the standard FTOOL \texttt{barycorr} and the DE-200 Solar System ephemeris \citep{solarDE200}. As the orbit of the binary system is unknown we could not correct for black hole's movement. We use the same source regions as for the spectral extraction and extract light curves with a time resolution of 10\,s between 3--10\,keV and 10--79\,keV. 

To measure the power spectral density (PSD) we use the method described in \citet{bachetti15a} to account for the variable dead-time of the \nustar detectors, which has typical values around 2.5\,ms. In this method, the PSD is calculated using the cross-power spectral density (CPSD) between FPMA and FPMB, thus removing the contribution of the (independent) dead-time and therefore allowing us to accurately measure the PSD to frequencies above 1\,Hz. We extracted the CPSD  using MaLTPyNT\footnote{\url{https://bitbucket.org/mbachett/maltpynt}}. 

To calculate  the CPSD we multiply the Fourier transformation of the light curve of FPMA with the complex conjugated Fourier transformation of the light curve of FPMB. The real part of the resulting complex quantity is the CPSD, describing power  in phase in both modules. As the deadtime is not correlated between FPMA and B and consequently not in phase, its power is removed from the CPSD.  For details about the method see \citet{bachetti15a}.

We calculated the CPSD from light curves of 64\,s length, extracted with a time resolution of $2^{-8}$\,s in the energy range  3--78\,keV based on the combined the event files from mode 01 and 06 and used a 150\asec source region.
The CPSD were averaged together for each satellite revolution, and geometrically rebinned with a factor 1.03. Uncertainties were propagated in quadrature from the unbinned data.

All CPSDs are presented in root mean square (RMS) or Miyamoto normalization \citep{belloni90a, miyamoto91a}.

\section{Analysis}
\label{sec:analysis}

\subsection{Light curve}
\label{susec:lc}
\begin{figure}
\includegraphics[width=0.95\columnwidth]{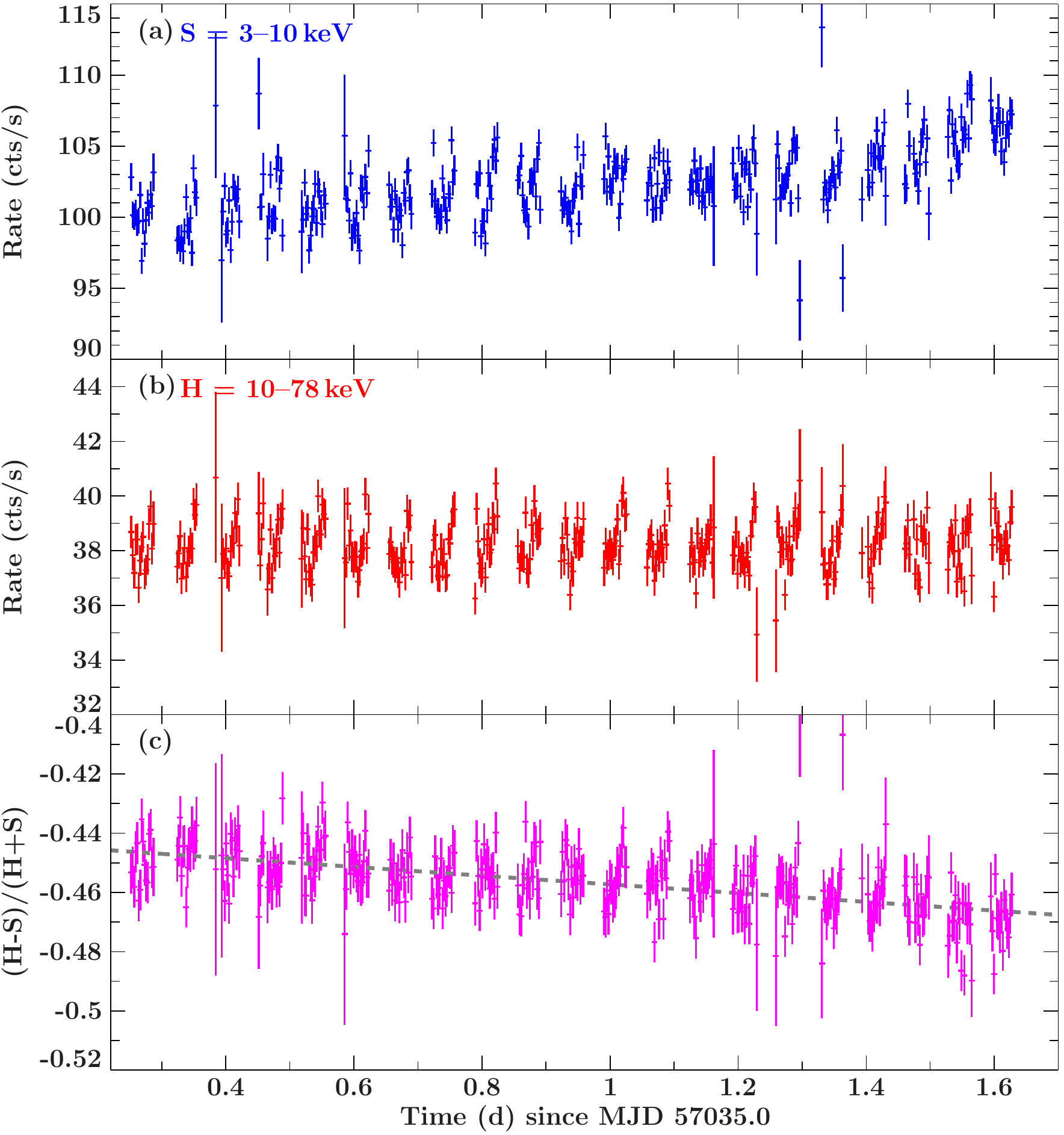}
\caption{\textit{(a)} \nustar light curve for the 3--10\,keV energy band (S); \textit{(b)} light curve for the 10--78\,keV energy band (H)  \textit{(c)} hardness ratio between the two light curves. The dashed line shows the best fit linear fit to the hardness ratio. The data are shown for FPMA only and are binned with a resolution of 200\,s.}
\label{fig:lc}
\end{figure}

To investigate the variability during the observation we first extracted  light curves in the 3--10\,keV (S) and the 10--78\,keV (H) energy bands with 200\,s time resolution. The light curves are shown in Figure~\ref{fig:lc} together with the hardness ratio (HR) calculated as HR = (H$-$S)/(H+S). The source softens steadily over the course of the observation, while the hard flux remains constant. 

This softening indicates a significant variation in the observed X-ray spectrum and thereby a change in the underlying physical conditions. In order to avoid combining data with intrinsically different spectra as much as possible, we decided to split the data into segments each covering one revolution of \nustar around Earth, i.e., we split the data at the gaps due to eclipses in the light curve. Thanks to \nustar's high sensitivity this still allows us to extract high signal-to-noise (\snr) energy  spectra and PSD.

\subsection{Time-resolved energy spectroscopy}
\label{susec:tres_spec}

After splitting the data according to each revolution  of \nustar, we obtained 21 energy spectra with about 2.7\,ks exposure each. We rebinned each spectrum to about 300 bins, with a \snr ratio of 24 at the lowest energies decreasing to a \snr ratio of 8 by 50\,keV. In the following analysis each spectrum was fitted individually, unless otherwise noted.
Following typical procedure and for comparability with other analyses we start modeling the data with simpler models and increase the complexity of the model step-by-step.

We first describe the data   with is an  absorbed power-law with an exponential turn-over (\texttt{cutoffpl}) and an additional reflection component, described by the \texttt{xillver} model \citep{garcia10a}. The \texttt{xillver} model self-consistently describes the reflection spectrum of an optically thick, geometrically thin accretion disk, including the \feka line as well as the Compton-hump at high energies. 

The continuum was modified at low energies with the absorption model \texttt{phabs} using abundances by \citet{wilms00a} and cross-sections by \citet{verner96a}. We set the equivalent hydrogen column to $5\times10^{21}$\,cm$^{-2}$, which is the expected Galactic absorption column along the line of sight towards \gx. The column could not be constrained  with the \nustar data and we therefore kept it fixed.  We also allowed for a cross calibration constant between the two \nustar focal plane modules, setting it to 1 for FPMA.

This model (model~1) gives us 7 free parameters per spectrum: the continuum flux, the photon index $\Gamma$, the cutoff energy $E_\text{cut}$, the reflection flux $A_\text{refl}$, the iron abundance $A_\text{Fe}$, the ionization parameter $\xi$, and the cross-calibration constant. 
The abundances in the \texttt{xillver} model are relative to the solar values found by \citet{grevesse98a}, i.e., an iron abundance of 1 corresponds to  $2.5\times10^{-5}$ Fe-atoms per the H-atom.

Model~1 results in an average $\redchi = 1.12$  ($\chi^2=13552$) for 578 degrees of freedom (d.o.f.) for each of the 21 spectra, i.e., 12138 total d.o.f. Note that we give the combined $\redchi$ as an indicator of the overall quality of the model, but that all spectra were fitted individually. The average X-ray luminosity (between 5--50\,keV) over the complete observation  is $L_x\approx6.35\times10^{37}$\,erg\,s$^{-1}$ for a distance of  8.4\,kpc, i.e., $\sim$$5.5\%\,\ledd$, assuming a 9\,\msun black hole. The average iron abundance is found to be $A_\text{Fe}=1.9\pm0.3$, where the uncertainty indicates the standard deviation  over the 21 data sets.

\begin{figure}
\includegraphics[width=0.95\columnwidth]{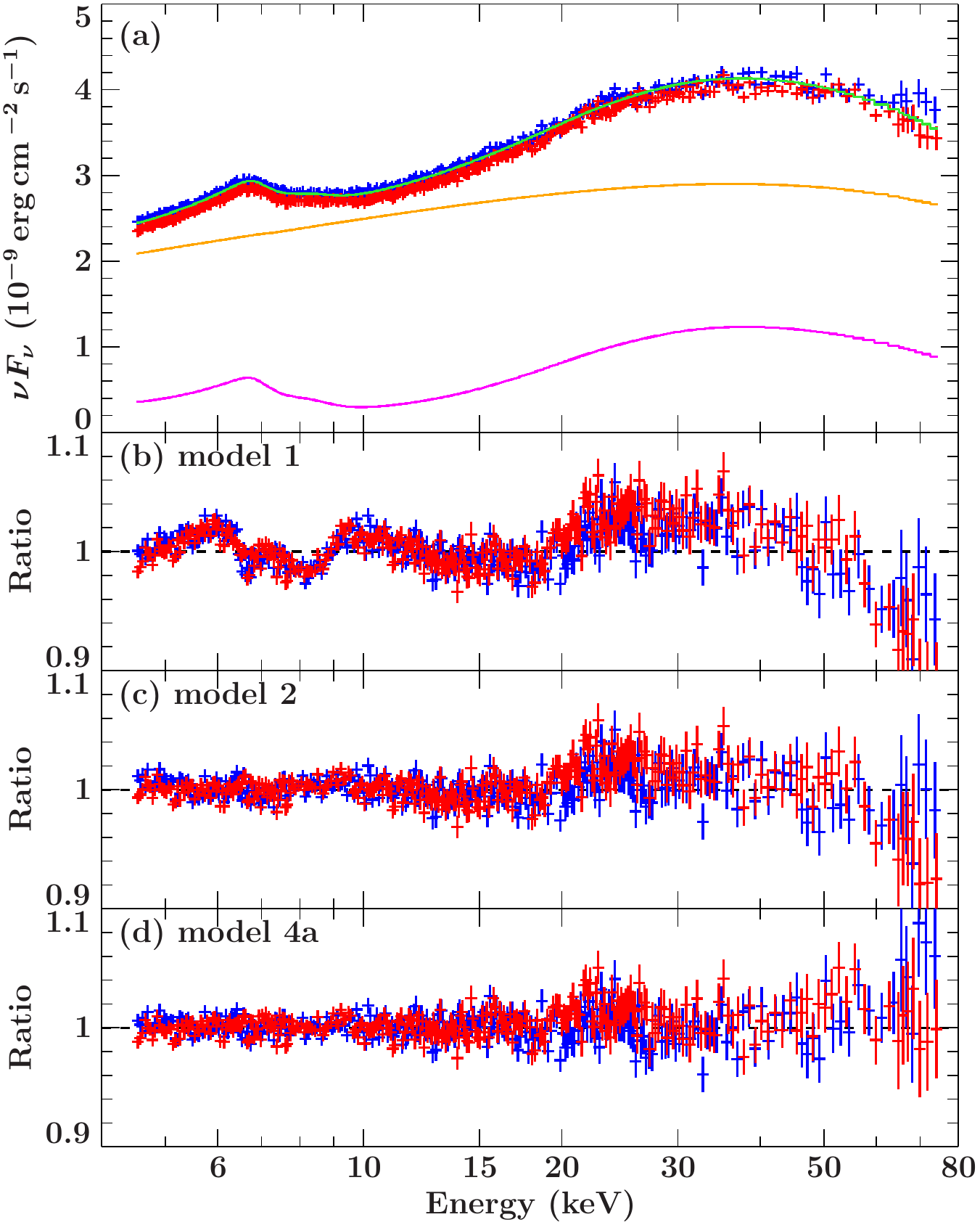}
\caption{\textit{(a)} Average of all 21 individually unfolded spectra, FPMA is shown in red, FPMB in blue. Superimposed in green is the weighted average of the best-fit models 4a for FPMA, thereby taking the variability of the spectra into account \citep[see][for a description of the method]{kuehnel16a}. The corresponding power-law and reflection components are shown in orange and  magenta, respectively. \textit{(b)} Residuals in terms of data-to-model-ratio for model 1, a reflection spectrum without relativistic smearing. Residuals are calculated for each spectrum separately and their average is shown. \textit{(c)} Residuals for model 2 which includes relativistic smearing. Residuals for mode 3 are not show but are very similar to model 2. \textit{(d)} Residuals for model 3a, in which the photon indices of the continuum and the reflector are independent. For details about the fitting process see text.}
\label{fig:spectra}
\end{figure}

 The average $\redchi$ value of model~1 is statistically not quite acceptable and an inspection of the combined residuals of all 21 spectra reveals strong structure around the \feka line as well as at the highest energies (Figure~\ref{fig:spectra}b). In  each individual spectrum the model seems to describe the data well (hence the relatively good $\redchi$ value), but the combination of the residuals of all spectra reveals strong structure. 

To improve the model we added a relativistic convolution model \citep[\texttt{relconv},][]{dauser10a}, broadening the reflection features as expected in the vicinity of the black hole (model 2). As in F15, we fixed the spin of the black hole to $a=0.93$, as found by \citet{miller08a}. \citet{parker16a} found a slightly higher spin, $a=0.95_{-0.08}^{+0.02}$, and we carefully checked that this increased value does not influence our results significantly.
We  kept the emissivity index of the accretion disk fixed at $q=3$, as expected for a standard Shakura-Sunyaev accretion disk \citep{reynolds97b} with a large coronal height, and the outer disk radius fixed at $r_\text{out}=400\,r_g$. 

The addition of relativistic blurring improved the fit drastically to an average $\redchi=1.03$ ($\chi^2=12422$)  for 576  d.o.f. per spectrum (the additional fit parameters being the the inclination $i$ and the inner accretion disk radius $R_\text{in}$). This is an improvement of $\Delta\chi^2 = 1130$ for 42 fewer d.o.f., clearly preferring this model over model 1 according to the sample-corrected Akaike Information Criterion \citep[AIC,][]{akaike74a}, with $\Delta\text{AIC}_{1,2} =1043$ (corresponding to a likelihood $\ll10^{-15}$) between model 1 and model 2.
The average inclination is $i=33\pm3^\circ$ and the average iron abundance $A_\text{Fe} = 5.3\pm1.3$. 
Allowing for a free spin $a$ resulted only in a marginal improvement of the statistical quality of the fit ($\Delta\chi^2 =12.9$) and in consistently high values of the spin in all 21 data-sets, validating our approach of fixing the spin to the literature value.
 Inspection of the combined residuals of model~2 again revealed  strong deviations at high energies, as shown in Figure~\ref{fig:spectra}c. 


In F15 we showed that there is evidence that the reflector sees a different continuum than the observed primary continuum during the hard state of \gx, and this was confirmed by \citet{parker16a}. This configuration can be crudely modeled by allowing the photon-indices of the continuum and the reflection model to be different (model 3a). With this approach, we obtain a very good fit with $\redchi= 1.007$ $(\chi^2=12158)$ for 575 d.o.f. per spectrum. We find   $\Delta\text{AIC}_{2,3}=220$, between models 2 and 3a,  indicating that the latter is clearly preferred with a chance likelihood  $\ll 10^{-12}$, which is confirmed by an F-test. 
The residuals to  model~3a are flat, as shown in Fig.~\ref{fig:spectra}d. Small deviations on the order of 2--3\% around 25\,keV can be attributed to calibration uncertainties \citep{nustarcalib}. These can be reduced by adding an ad-hoc Gaussian component at $\sim$22\,keV, but this addition does not change any of the fit parameters significantly. As we already obtain a very good fit in terms of $\redchi$ and to keep the model as simple as possible, we chose not to include this  component.

\begin{figure}
\includegraphics[width=0.95\columnwidth]{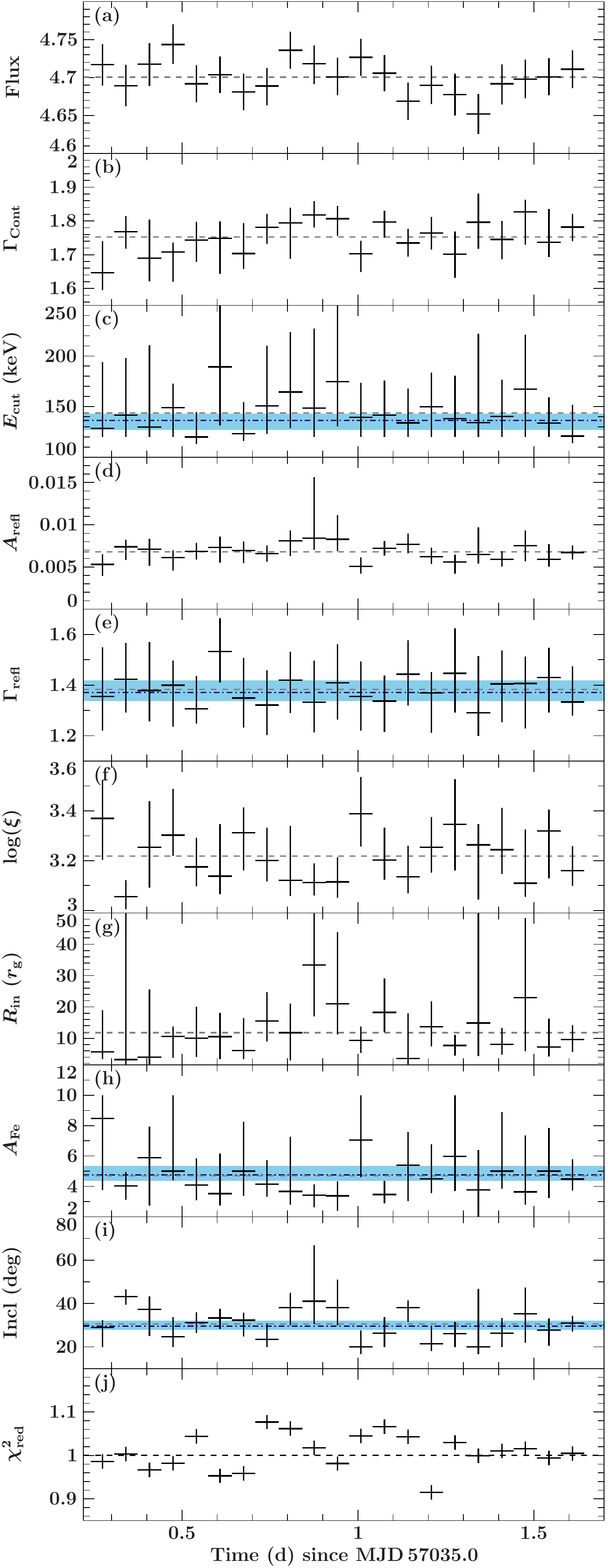}
\caption{Results of the spectral fits on a revolution by revolution basis for model 3a. \textit{(a)} X-ray flux between 5--50\,keV in keV\,s$^{-1}$\,cm$^{-2}$, \textit{(b)} photon index of the continuum, \textit{(c)} cutoff energy in keV, \textit{(d)} normalization of the reflection spectrum, \textit{(e)} photon index of the input spectrum to the reflection, \textit{(f)} ionization parameter in erg\,cm\,s$^{-1}$, 
\textit{(g)} inner accretion disk radius in gravitational radii, \textit{(h)}  iron abundance, \textit{(i)} inclination in degrees, and \text{(j)} reduced $\chi^2$ fit statistic. The gray dashed lines in each panel indicate the respective average value over all 21 spectra and the blue dot-dashed lines the results from the simultaneous fit, together with its 90\% uncertainty shaded in light blue. For details see text.}
\label{fig:specres}
\end{figure}
\begin{figure}
\includegraphics[width=0.95\columnwidth]{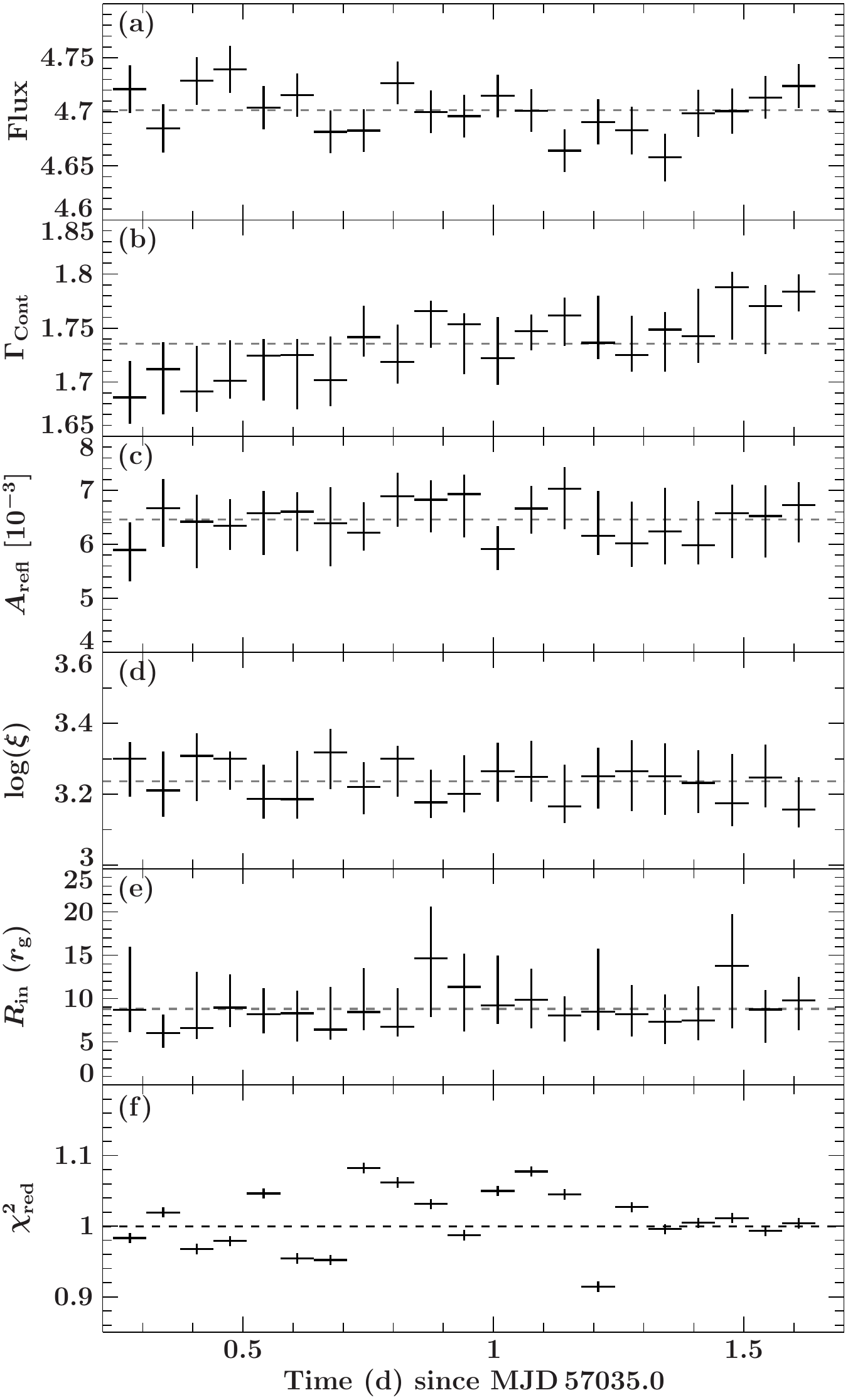}
\caption{Same as Figure~\ref{fig:specres}, but for model 4 in which the global parameters were allowed to vary only with the 90\% uncertainties shown in Table~\ref{tab:simfitres} and are therefore not plotted.  \textit{(a)} X-ray flux between 5--50\,keV in keV\,s$^{-1}$\,cm$^{-2}$, \textit{(b)} photon index of the continuum, \textit{(c)} normalization of the reflection spectrum, \textit{(d)} ionization parameter in erg\,cm\,s$^{-1}$,  \textit{(e)} inner accretion disk radius in gravitational radii, and \textit{(f)} reduced $\chi^2$ fit statistic. 
The  dashed lines in each panel indicate the respective average value over all 21 spectra.}
\label{fig:specres_parofsimfit}
\end{figure}

Another way of approximating a complex coronal configuration is to allow the cutoff energy of the primary continuum and the input spectrum to the reflector to be different (model 3b). This setup might mimic different temperatures in an elongated corona. Using this approach, we find a slightly worse fit than with independent photon indices and obtain $\redchi=1.013~(\chi^2=12232)$ for 575 d.o.f.  per spectrum. We obtain qualitatively similar results, with the input spectrum to the reflector being much harder (hotter) than the observed continuum. However, the cutoff energy of the reflector often pegs at the upper limit of 1\,MeV. As this model also gives a statistically slightly worse fit than model 3a ($\Delta\chi^2=74$ for the same number of d.o.f.), which allowed for independent photon-indices, we will build on model 3a for the remainder of this paper. The physical implications (e.g., the values of the inner radii) do not differ significantly between these two models.

\subsubsection{Simultaneous Fits}
In the previous models, we kept all parameters independent between each spectrum. 
We can assume, however, that  certain parameters do not change over the course of the observation, like the inclination $i$ and the iron abundance $A_\text{Fe}$\footnote{Or the spin $a$, which has been held fixed.}, as long as the accretion disk is not strongly warped and precessing \citep[as postulated for, e.g., Cyg~X-1,][]{tomsick14a}.
To implement such constant parameters in our model we need to perform a simultaneous fit of all 21 spectra.

Figure~\ref{fig:specres} shows the evolution of all fit parameters as a function of time in model 3a. 
While we know from the change in hardness that a significant evolution in the spectrum is present, the relatively large uncertainties on all parameters in model~3a render them insensitive to these changes. To quantify the spectral changes, we therefore have to choose parameters to which we ascribe the observed change in hardness, while we require others to not change over the course of the observation.

We select the photon index of the continuum  $\Gamma_\text{cont}$ as the main driver of the hardness changes, and tie the cutoff energy $E_\text{cut}$ and the photon index of the reflector $\Gamma_\text{refl}$ between all spectra in the simultaneous fits. This selection is somewhat arbitrary, but results in a very good description of the data. We note that if we instead allow the photon index of the reflector, $\Gamma_\text{refl}$, to vary, and tie the photon index of the continuum, $\Gamma_\text{cont}$, across all spectra, we obtain very similar correlations and implications. We will base our discussion on a variable photon index of the continuum with the understanding that this is not necessarily the correct physical parameter, but just a convenient parameterization of the hardness.

 For the simultaneous fit, we  have  the parameters  $i$, $A_\text{Fe}$, $E_\text{cut}$, $\Gamma_\text{refl}$, and the cross-calibration constant for FPMB, $CC_\text{FPMB}$ tied across all  spectra. These are ``global'' parameters as  defined by \citet{kuehnel16a}. 
This simultaneous fit provided a statistically very good description of the data, with $\redchi=1.009~(\chi^2$=12290 for 12175 d.o.f.). 

However, as this fit is highly complex and a single evaluation of the spectrum takes almost 2\,s on a typical workstation, it was not feasible to calculate uncertainties for all 110 free parameters. Instead, we only calculated uncertainties for the parameters tied across all data-sets (Table~\ref{tab:simfitres}). In a second step, we go back to  modeling the 21 spectra individually, but constrain the global parameters to their 90\% uncertainty interval found in the simultaneous fit (model~4). 
This approach ensures that the values of the global parameters  are similar across all data-sets, while still capturing their statistical variability and allows us to obtain a realistic estimate for uncertainties of the local parameters. 

\begin{table}
\caption{Parameters and uncertainties of the best-fit model by fitting all 21 spectra simultaneously.}
\begin{center}
\begin{tabular}{c|c}
Parameter & Value   \\\hline
$E_\text{cut}$ (keV) & $136^{+6}_{-9}$ \\
$\Gamma_\text{refl}$ &  $1.37^{+0.05}_{-0.04}$  \\
$A_\text{Fe}$ & $4.8^{+0.6}_{-0.4}$  \\
$i$ (deg)& $29.7^{+2.2}_{-1.5}$  \\
$CC_\text{FPMB}$ &  $1.0299^{+0.0013}_{-0.0012}$   \\\hline
\end{tabular}
\end{center}
\label{tab:simfitres}
\end{table}%
%

Model~4 provided a  good fit to the data, with $\redchi=1.012~(\chi^2$=12221 with 575 d.o.f. per spectrum). The results of the local parameters are shown in Figure~\ref{fig:specres_parofsimfit}. The global parameters do not show a trend as function of time, and their uncertainties encompass the complete allowed range in each individual spectrum, i.e., they follow the blue bars shown in Figure~\ref{fig:specres}.
As is  obvious in a comparison  of Figures~\ref{fig:specres} and  \ref{fig:specres_parofsimfit}, the local parameters are  better constrained in model~4, in particular the photon index $\Gamma_\text{cont}$ and the inner radius of the accretion disk, $R_\text{in}$. As expected from our model setup, the only parameter that shows a significant evolution as a function of time is $\Gamma_\text{cont}$; all other parameters are constant within their uncertainties. The photon index $\Gamma_\text{cont}$, however, shows a strong steepening over the course of the observation, in step with the observed softening shown in Figure~\ref{fig:lc}. 
The $\redchi$-values of this model vary around $1$, in a manner fully consistent with a $\chi^2$-distribution with 580 d.o.f. according to the KS-test.

To test the dependence of the values of the inner radius on the assumed geometry, we also tested the lamppost geometry using the  \texttt{relxilllp} model \citep{dauser14a}, which assumes a point-like corona on the spin-axis above the black hole. We find a statistically comparable good fit, using the same approach of allowing the global parameters  to only vary within their 90\% uncertainties found in a simultaneous fit.  
 All parameters, in particular the photon index and the inner radius, follow the results from model~4 closely.
The similarity between the two models indicates that in our data the exact geometry and emissivity profile of the accretion disk do not significantly influence the results.

\subsection{Periodogram}
\label{susec:psd}

\begin{figure}
\includegraphics[width=0.95\columnwidth]{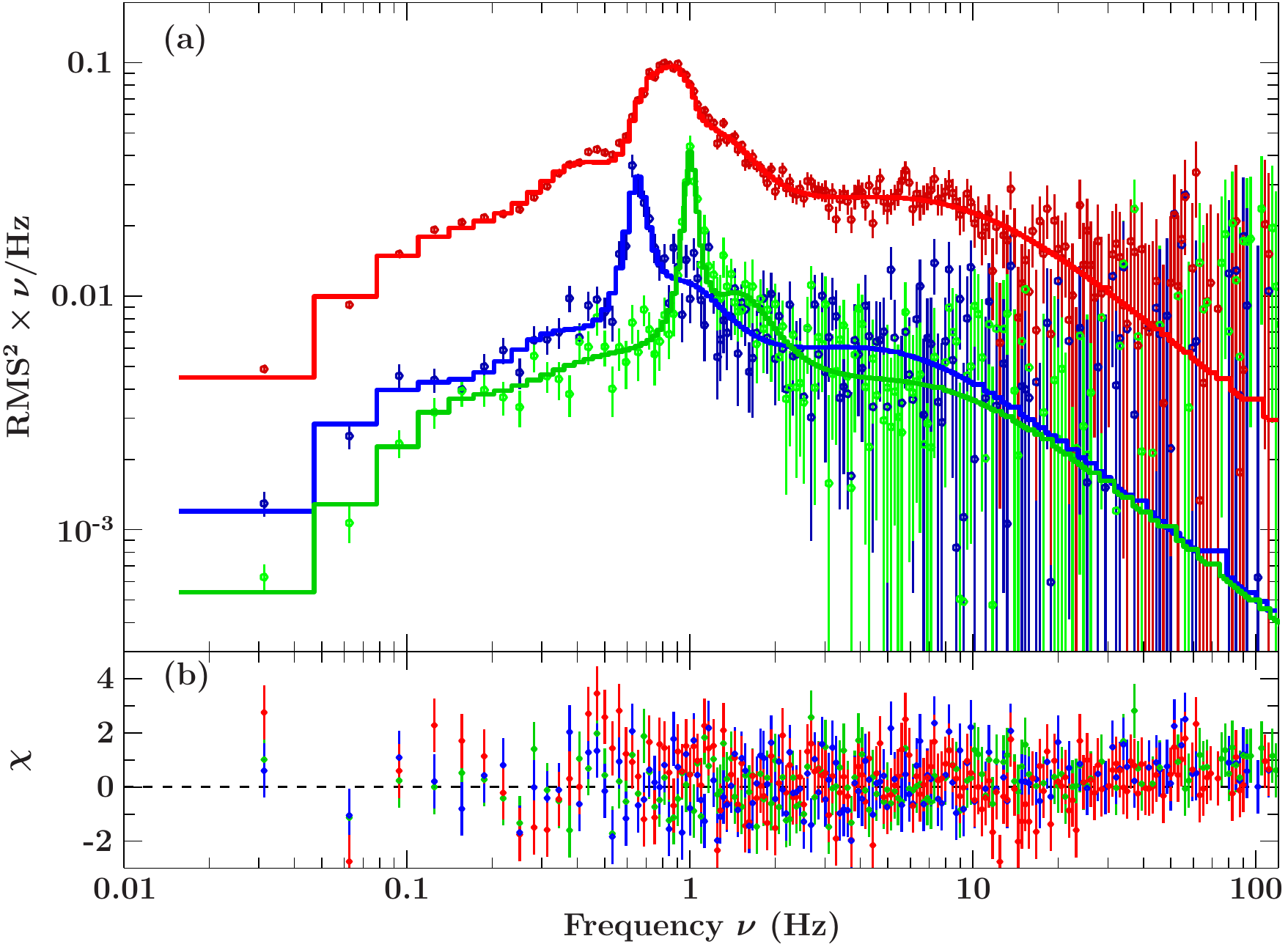}
\caption{\textit{(a)} Average CPSD in red and the CPSD of the first and last revolution in blue and green, respectively, together with their best-fit models. The average CPSD is multiplied by a factor of 5 for visual clarity. The time-dependence of the QPO frequency around 0.8\,Hz  is clearly visible. \textit{(b)} Residuals to the best-fit models in terms of $\chi^2$.}
\label{fig:pds}
\end{figure}

In the same way as for the time-resolved energy spectra, we analyzed the CPSD on a revolution-by-revolution basis. All 21 CPSD show a very prominent type-C QPO around 0.8\,Hz  \citep[for a definition of QPO types, see, e.g., ][]{casella05a}. We modeled each CPSD with two zero-centered Lorentzians to describe the band-limited noise, a Lorentzian for the type-C QPO and one Lorentzian for the sub- and super-harmonic (at $\frac12$ and $\frac32$ the frequency of the QPO) each. This model provides a very good fit to the data, with an average $\redchi=1.01$ for 183 d.o.f. Figure~\ref{fig:pds} shows the average CPSD and best-fit model in red. The QPO around 0.8\,Hz appears broad, as it is a superposition of individual QPOs in each individual CPSD.

\begin{figure}
\includegraphics[width=0.95\columnwidth]{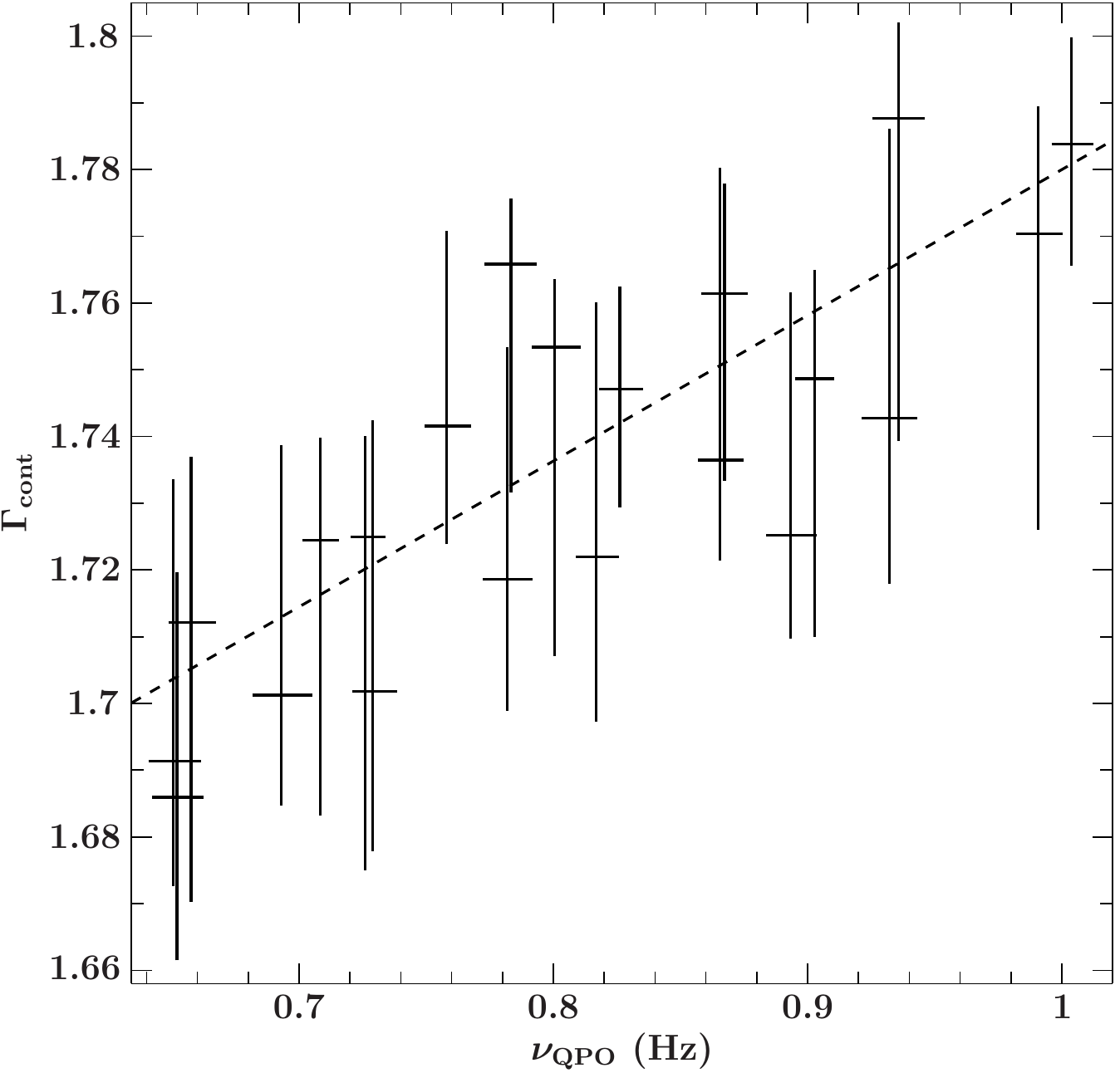}
\caption{Correlation between the best-fit QPO frequency \nuqpo and the photon index of the power-law continuum $\Gamma_\text{cont}$ from model~4. Superimposed is the best-fit linear correlation with a slope of $0.22\pm0.11$.}
\label{fig:qpo2gamma}
\end{figure}

By looking at the QPO frequency as function of time, we find that it is significantly increasing, rising from $\sim$0.65\,Hz at the beginning of the observation to $\sim$1.0\,Hz at the end (Figure~\ref{fig:pds}, in blue and green respectively). This change is in step with the softening of the photon index $\Gamma_\text{cont}$, as shown in Figure~\ref{fig:qpo2gamma}, where we plot the QPO frequency versus $\Gamma_\text{cont}$ together with a linear correlation with a slope of $b=0.22 \pm 0.11$.
This correlation is often observed for type-C QPOs, in \gx as well as in other sources \citep[e.g.][]{vignarca03a, motta11a, stiele13a}.

We do not find a significant correlation between the inner accretion disk radius and the QPO frequency, with a Pearson correlation coefficient of 0.16 and a linear slope of $b_\text{rin}=6\pm 12$. The uncertainty is dominated by the larger uncertainty on the inner radius, which cannot be constrained better than $\approx3.5\,r_g$.

To investigate the dependence of the QPO frequency on energy we extracted CPSDs in the energy bands 3--5\,keV, 5--7.5\,keV, 7.5--12\,keV, and 12--79\,keV. We do not see any significant change in frequency as a function of energy, and all four energy bands give the same time dependence. While type-C QPOs show a strong energy dependence in some sources if they are located at higher frequencies \citep[$\gtrsim4$\,Hz for H1743$-$322,][]{li13a}, at lower frequencies they are energy independent \citep[e.g., in XTE\,J1550$-$564,][]{cui99a}, in agreement with our results. 

We carefully searched for evidence of QPOs at higher frequencies, but did not find a significant signal. For that search we also analyzed the time averaged CPSD to obtain a better signal at high frequencies. Here, we find a slight improvement of the statistical quality of the fit when including a QPO at around 55\,Hz. However, by using a Monte Carlo approach to test its significance and simulating 500 CPSDs from the best-fit model, scattered around within the uncertainties of the measured CPSD, we find that a similar improvement in $\chi^2$ can occur by chance with $>5\%$ probability. We therefore conclude that this feature is not significantly detected.

\section{Discussion}
\label{sec:summ}

We have presented  time-resolved spectral and timing analysis of \nustar data of \gx, taken during a hard intermediate state in early 2015. The time-resolved approach was necessary as the source was softening significantly over the course of the observation and at the same time the prominent type-C QPO increased in frequency. From the energy spectra we concluded that the accretion disk is truncated around 9\,$r_g$, independent of the assumed geometry. This value is larger than the ISCO for a spin of $a\approx0.95$ ($r_\text{ISCO}=1.9\,r_g$) or $a\approx0.93$ ($r_\text{ISCO}=2.1\,r_g$), as measured by \citet{parker16a} and \citet{miller08a}, respectively. We note that using the relativistic effects to constrain the inner radius of the accretion disk  still has systematic uncertainties, as all available geometries are simple approximations, as discussed in F15. Independent of these assumptions, the data show that within the uncertainties, the inner radius is not changing over the course of the observation, despite the observed changes in spectral hardness (Figure~\ref{fig:specres_parofsimfit}\textit{e}).

Our best-fit model assumes that the power-law continuum incident on the reflector is harder than that observed as underlying continuum. This configuration was invoked in F15 for the hard state and later confirmed by \citet{parker16a} in a soft state observation, and indicates a complex geometry of the corona. Using this geometry, F15 found that the iron abundance is significantly reduced compared to other models. This is not the case in the observation presented here, and all models require an iron abundance around 4--5 solar, similar to the values found by \citet{parker16a} and \citet{garcia15a}, using \nustar and \xte data, respectively. 

\subsection{Truncation of the inner accretion disk}
Using relativistic reflection models, we find that the inner radius of the accretion disk is truncated significantly outside the ISCO, at around 9\,$r_g$. This result is independent of the assumed geometry, either a standard disk emissivity profile with an index of $q=3$, or a lamppost geometry.
This amount of truncation is similar to the one found by \citet{tamura12a}, assuming a 9\,\msun black hole.

This result indicates that the accretion disk does not move all the way to the ISCO before the source has entered the soft state. This seems to be at odds with results finding the accretion disk to be at the ISCO in bright hard states \citep[e.g.,][]{miller15a}. On the other hand, it might simply show that different outbursts behave differently. It is in theory also possible that the truncation is not a strict function of luminosity, and truncation appears again during the HIMS, given its slightly different radiation balance at the accretion disk/corona boundary (e.g., a corona with a steeper photon index and therefore likely higher optical thickness).

Finding a truncated accretion disk in the HIMS nonetheless indicates that the geometry and physical conditions in this state are very different than in the soft state, e.g., that the inner parts of the accretion disk are still replaced with an optically thin flow, like an ADAF. During the switch to the soft state, which happens on a very short time-scale, the accretion disk has then to move all the way to the ISCO and significantly heat up. As can be seen in Fig.~\ref{fig:batlc}, the source is in the proper soft state only 3--4 days after the \nustar observation.

\subsection{QPOs and the inner accretion disk radius}
\label{susec:qpoorig}
Despite being an ubiquitous feature of black hole binaries, the physical origins of QPOs are still highly uncertain. While QPOs come in different shapes \citep[see, e.g.][]{casella04a}, the most often observed one is the type-C QPO, i.e., a strong, narrow feature in the PSD on top of flat-top noise dominated continuum  \citep{remillard02a}. 
Different theories have been put forward as to their origin, with one of the most prevalent being that the QPO is produced by Lense-Thirring precession of the innermost parts of the accretion flow \citep[e.g.][]{stella98a,stella99a,ingram09a}.

This model, known as the relativistic precession model (RPM), predicts three QPOs to be observed, all of which are produced by the same accretion flow. The lowest frequency QPO is identified with the nodal precession of the accretion flow due to Lense-Thirring precession and observable in black hole binaries as a type-C QPO around 0.5--5\,Hz \citep{stella99a}. If all three QPOs are observed \citep[as in, e.g., GRO~J1655$-$40,][]{motta14a}, the RPM allows for a measurement of the mass and spin of the black hole, as well as the inner accretion disk radius, independent of results or assumptions from fitting the energy spectra.

\begin{figure}
\includegraphics[width=0.95\columnwidth]{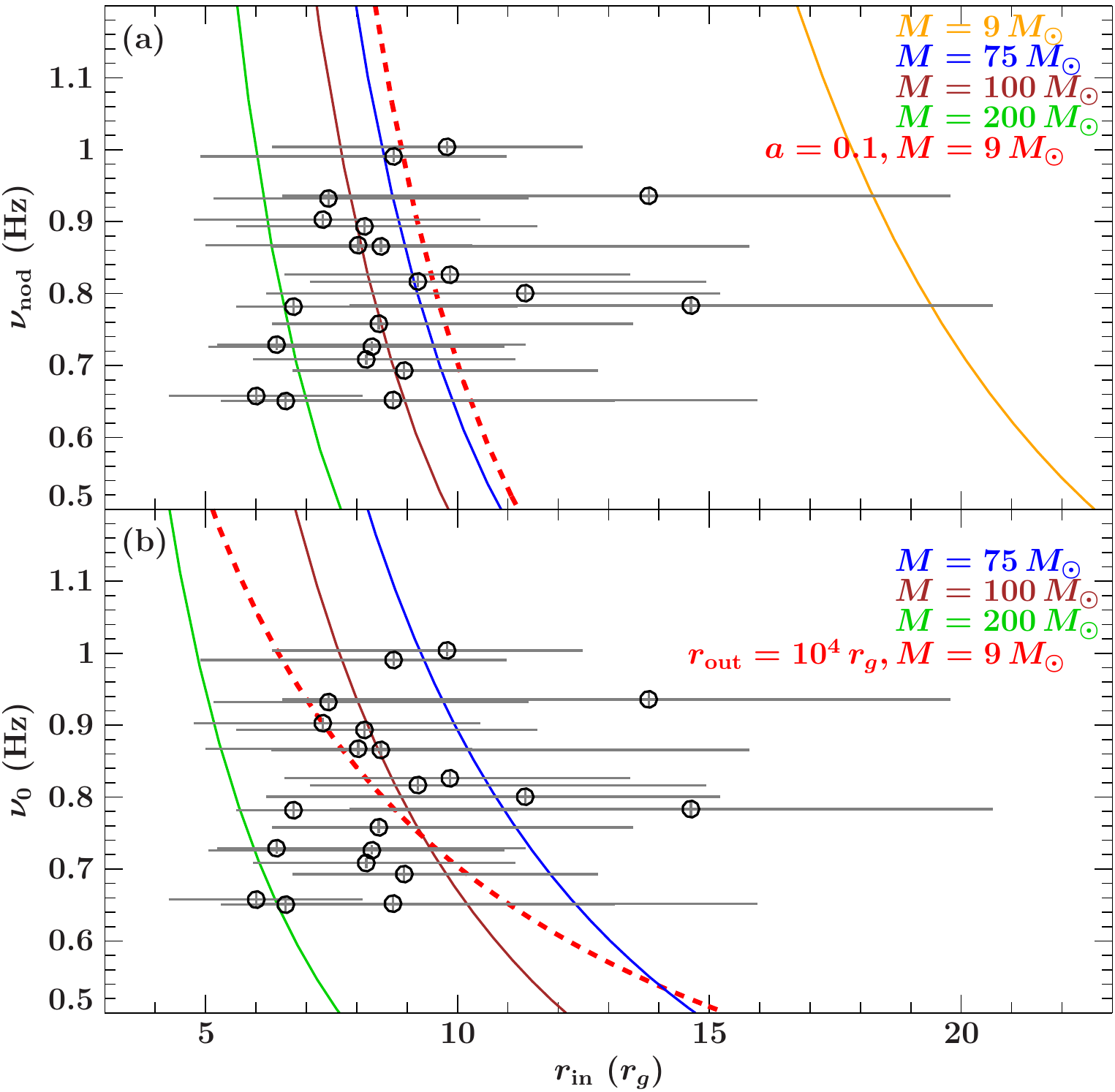}
\caption{\textit{(a)} Nodal precession frequency $\nu_\text{nod}$ in the RPM model, which can be identified with the type-C QPO frequency, as function of inner radius $r_\text{in}$ for a given black hole mass and an assumed spin of $a=0.95$. The solid colored lines show the correlation for different assumed black hole masses. The dashed red line is calculated for a spin of $a=0.1$ and a mass $M=9\,\msun$. The black data points and their uncertainties show the measured QPO frequencies and inner radii (model~4) in the 21 spectra. \textit{(b)} Same as \textit{(a)}, but for the disk oscillation frequency $\nu_0$ in the GDM model. The 9\,\msun line lies outside the plot to the right  at larger inner radii. The red dashed line assumes an outer radius of $10^4\,r_g$ and a  mass $M=9\,\msun$. }
\label{fig:nod2rin}
\end{figure}

In \gx, we only observe the low-frequency QPO, which can be identified with the nodal precession frequency $\nu_\text{nod}$. If we include measurements of the spin and inner radius from reflection modeling, we can investigate possible black hole masses. We follow the calculations presented in \citet{ingram14a}, and assume a spin of $a=0.95$ \citep{parker16a}. Using the measurements of the inner radius of model 4, Figure~\ref{fig:nod2rin}\textit{a} shows the correlation between $\nu_\text{nod}$ and $r_\text{in}$ for different masses of the black hole, as predicted by the RPM. As can be seen, our measured values are incompatible with a black hole mass of $\sim$9\,\msun and a fast spinning black hole. Instead they seem to indicate a very massive black hole around 75--100\,\msun, which is too large for typical  X-ray binaries. Within the RPM the sampled range of QPO frequencies results only in a very small variation of the inner radius, clearly below the uncertainties of the measured values.

As shown in Figure~\ref{fig:nod2rin}\textit{a} we can consolidate the measurements of the inner radius and QPO frequency with a black hole mass of 9\,\msun when assuming a very low spin of $a=0.1$. Such a low spin is in stark contrast to all relativistic reflection modeling where a broad \feka line is observed \citep{reis08a, miller08a,  ludlam15a, garcia15a, parker16a} and also ruled out by disk continuum fitting for inclinations $i>20\deg$ \citep{kolehmainen10a}.

Another model to explain the origin of low-frequency QPOs was presented by \citet{titarchuk00a}. These authors assume that the QPO is produced by global disk mode (GDM) oscillations, i.e., by the oscillations from a disk off-set from the equatorial plane around a compact object without the need of invoking relativistic effects. In this theory the total mass of the disk plays an important role, which is strongly influenced by our choice of  the outer radius when assuming a standard Shakura-Sunyaev disk. Using an outer radius of 400\,$r_g$, and the measured inner radii and QPO frequency, we find that the data are best described with a 100\,\msun black hole (Figure~\ref{fig:nod2rin}\textit{b}).  

However, as the outer radius is not known and cannot be constrained in our spectral fits, we can also find an acceptable solution for a 9\,\msun black hole with an outer accretion disk radius of $10^4\,r_g$. We note, however, that the predicted correlation between $r_\text{in}$ and $\nu_0$ in that case seems somewhat flatter than what we measure, as shown in Figure~\ref{fig:nod2rin}\textit{b}. 

\subsection{QPOs and the photon index}

\citet{motta11a} and \citet{stiele13a} present a detailed study of the correlation between QPO frequency and spectral parameters in \gx using \xte data of outbursts between 2002--2010. They find a clear correlation between QPO frequency and photon index, which, for hard spectra and frequencies around 1\,Hz, is linear. While their data sample a large range in both $\Gamma$ and $\nu_\text{QPO}$, the sampling during the transitional state around $\Gamma\approx1.7$ is very sparse. This makes a comparison to our data not feasible, but places the unique state we sampled with \nustar in context.

\citet{shaposhnikov09a} use a similar selection of \xte data to compare the correlation of \gx to the one observed in other sources, e.g., XTE~J1650$-$500 and Cyg~X-1. By following the idea that the differences in the observed QPO-$\Gamma$ correlations between different sources can be reduced to differences in the mass of the compact object \citep{titarchuk04a}, they estimate a black hole mass for \gx of $12.3\pm1.4\,\msun$. This method requires the measurement of the correlation over a large range of QPO frequencies, in particular covering the transition regime, at which the correlation flattens. 

The \nustar data cover only a very small range of QPO frequencies, but assuming that the measured range is well below the transition frequency (which is, e.g., at $6.64\pm0.48$\,Hz for GRO~J1655$-$40, \citealt{shaposhnikov07a}), we can fit a linear function to the correlation. This gives a slope of $b=0.22\pm0.11$ (see Figure~\ref{fig:qpo2gamma}). Comparing this to  values for other sources presented by \citet{shaposhnikov07a}, we  estimate a mass of $\sim$10\,\msun for \gx from the scaling relation, in agreement with the mass measured by \citet{parker16a}. This value is dominated by systematic uncertainties given the limited QPO frequency range covered during the  \nustar observation and therefore  compatible with the mass found by \citet{shaposhnikov09a}.

\subsection{Summary}

By combining precise  spectral modeling with recent theories concerning the formation of QPOs in accreting black holes, we have shown that there are still many questions left to answer. Our data show that in both the relativistic precession model  as well as the global disk mode oscillation model, a black hole mass on the order of 100\,\msun seems to be necessary to explain the observed combination of QPO frequency and inner accretion disk radius. As such a massive black hole is highly unlikely to exist in \gx \citep{parker16a}, it shows that  further improvement of the measurements and QPO theory is necessary. We find a significantly truncated accretion disk ($r_\text{in}\approx4.4\,r_\text{ISCO}$ for $a=0.95$) but the measurement of its inner radius through reflection fitting is strongly dependent on emission geometry and inclination of the accretion disk. While we have shown that  the lamppost geometry gives similar inner radii as a standard emissivity profile, more complicated coronal geometries cannot be ruled out. The measurement of the photon index $\Gamma_\text{cont}$ is more reliable, for which we find a tight correlation to the observed QPO frequency. However, theoretical predictions based on physical models are currently lacking, due to the fact that the physics of the corona are still poorly understood. Further high quality observations similar to the ones presented here will allow us to increase our knowledge about the corona and its physics in the future.

\acknowledgments
We thank the anonymous referee for the  constructive and helpful comments.
We thank the \nustar schedulers and SOC, in particular Karl Forster, for making this observation possible.
We thank Javier Garc\'ia and Thomas Dauser for helpful discussions about the reflection models.
This work was supported under NASA Contract No. NNG08FD60C, and
made use of data from the {\it NuSTAR} mission, a project led by
the California Institute of Technology, managed by the Jet Propulsion
Laboratory, and funded by the National Aeronautics and Space
Administration. 
Support for this work was provided by NASA through the Smithsonian Astrophysical Observatory (SAO) contract SV3-73016 to MIT for Support of the Chandra X-Ray Center (CXC) and Science Instruments; CXC is operated by SAO for and on behalf of NASA under contract NAS8-03060.
We thank the {\it NuSTAR} Operations, Software and
Calibration teams for support with the execution and analysis of these observations. 
This research has made use of the {\it NuSTAR}
Data Analysis Software (NuSTARDAS) jointly developed by the ASI
Science Data Center (ASDC, Italy) and the California Institute of
Technology (USA). 
This research has made use of a collection of ISIS functions (ISISscripts) provided by ECAP/Remeis observatory and MIT (\url{http://www.sternwarte.uni-erlangen.de/isis/}).
We would like to thank John E. Davis for the \texttt{slxfig} module, which was used to produce all figures in this work. 
This research has made use of MAXI data provided by RIKEN, JAXA and the MAXI team.

{\it Facilities:} \facility{NuSTAR}


\end{document}